\documentclass[runningheads]{llncs}
\usepackage[english]{babel} 
\usepackage{ucs}
\usepackage[utf8x]{inputenc}
\usepackage[T1]{fontenc}
\usepackage{pslatex}
\usepackage{graphicx}
\usepackage{xspace}
\usepackage{float}
\usepackage{color}
\usepackage{alltt}
\usepackage{boxedminipage}

\usepackage{tikz}
\usetikzlibrary{matrix}

\usepackage{tikz}
\usetikzlibrary{shapes,arrows}
\usepackage{amsmath,bm,times}
\usepackage{zedbis}

\usepackage[pdftex]{thumbpdf} 
\usepackage[bookmarks=true,bookmarksnumbered=true,colorlinks=false,urlcolor=blue,citecolor=blue,linkcolor=blue]{hyperref}

\def\repHOME{/home/attiogbe-c/Nextcloud/attiogbe/RAP}

\def\repBIBLIO{\repHOME/IoT/FormalModel_IOT/InvariantArchi/.}

\title{Architectural Invariants and Correctness of IoT-based Systems}
\author{Christian Attiogb\'e and Jérôme Rocheteau}

\institute{  LS2N CNRS UMR 6004 - University of Nantes\\
  \email{\{christian.attiogbe, jerome.rocheteau\}@ls2n.fr}}

\begin{document}
\maketitle

\begin{abstract}
  Internet of Things applications impact more and more industrial areas such as smart manufacturing, smart health monitoring and home automation; physical objects or devices equipped with sensors and actuators are interconnected  and then controlled with software applications.
  Ensuring the correct construction, the well functioning  and  the  reliability of these applications constitute important issues for some of these applications which can be critical in case of dysfunction.
  We propose on the basis of the formal model of their common architectural properties,
  a generic framework for
  the formal modelling of IoT-based applications,
  the rigorous analysis of their consistency properties,
  their rigorous construction and evolution.
  Specific properties can be gradually added and checked. The proposed framework is then implemented and experimented using Event-B.
  We exploit the observation that the main requirements of the IoT-based physical architectures and control software are common to all IoT-based applications; this leaded us to the definition of the generic formal model together with invariant architectural and consistency properties. 
The proposed approach is generic, extensible, and can be profitably adapted to more general hybrid or cyber-physical systems. Our current implementation is independent of the formal model, it can  be achieved in various other formal analysis environments. 
\end{abstract}



\section{Introduction}
Internet of Things applications impact industrial areas such as smart manufacturing, smart vehicles, smart logistics and transpiration, smart farming, etc.
Therefore IoT-based applications constitute an important part of software being developed and interconnected around the world;
they will be active for years and will need maintenance, due to new requirements, constraints, standards, evolution of materials and communication protocols, user-interfaces, etc.
Fortunately, in some extents, these applications share  a well-established architectural structuring, reference models, some functional and non-functional properties  \cite{Guth2018_AnalysisIoTPlatformArchitectures,DBLP:journals/comsur/Al-FuqahaGMAA15,refModelIoTMiddleware2018}.

However, well-established engineering methods and techniques are still needed \cite{DBLP:journals/jsa/Sosa-ReynaTA18} to ensure that the applications are reliable, secure, scalable, well integrated, and extensible. 

In this context we are motivated by proposing methods and tools for mastering the modelling, the analysis, the development and the maintenance of such IoT-based applications. The challenges are that these applications can become rapidly complex because of their evolving heterogeneous environment. Indeed the environment of an IoT-based application comprises hardware items, physical devices, software, control techniques, communication protocols, network services. Moreover this environment is continuously changing.

In order to ensure the consistency and the well-functioning of an IoT-based application, the latter should integrate as a parameter, the complete description of the physical context that it controls. Therefore a global model can be built and analysed with respect to consistency and to the required specific properties.  We propose such a global formal modelling and the related analysis.

The contributions of this paper are manifold: we propose 
\textit{i)} a generic formal description of the physical architecture of an IoT-based system;
\textit{ii)} a formal description of the control application parametrised by its physical environment;
\textit{iii)}  the modelling and analysis of the invariant architectural properties of such IoT-based systems and the description of some specific properties. They are described so as to be customizable for other application cases. We design a generic framework using Event-B to support the full modelling and analysis approach. Moreover we develop a tool that generates systematically for a given IoT-based application described with a domain specific language, the specific Event-B parts that are used to instantiate the generic Event-B framework. 

The organisation of the article is as follows.
In Section \ref{section:background}, we introduce the background for understanding the used concepts.
Section \ref{section:modelling} is devoted to the generic modelling of IoT-based physical structure and control application;
In Section \ref{section:consistencyModel}, we deal with the invariant consistency properties, which are formalised and checked.
In Section \ref{section:checkingArchitecture}, we show how we have implemented our proposed generic formal model and analysis technique using Event-B; we compare our work to the related ones.
Finally Section \ref{section:conclusion} presents some perspectives and future work.

\section{Basic Concepts and Architecture Elements of IoT Systems}
\label{section:background}

Based on existing state-of-the-art references \cite{DBLP:journals/comsur/Al-FuqahaGMAA15,Guth2018_AnalysisIoTPlatformArchitectures,Survey_IOT_Archi_RAY_elsevier2018,mbaVerifFGCS2019} which account of IoT technologies, challenges, comparisons, and reference models, we consider the following main elements of IoT. 

A \textit{thing} is a physical  object (like a specific device, a domestic or industrial robot, a door, a light, a watering system, etc) equipped with:
\textit{i)} sensors that collect  and gather data from the environment (devices to control, devices under measurement, etc);
\textit{ii)} actuators that allow the control of the thing or allow the thing to act on its environment. Examples of actions are: switching on or off a light, opening or closing a door, increasing or decreasing the speed of an engine, launching a robot, etc.
Sensors are not always physically binded to the things; but data from sensors can be transferred over a network to reach the thing. Similarly  actuators can be linked to the thing via a dedicated network.

An IoT-based application is a software built on top of the physical infrastructure made of one or several things. Such applications made of services, are used for monitoring or controlling various devices or processes.
An overview of the architecture of a control application is depicted in Fig. \ref{figure:iot-appli} where we can distinguish:
\textit{i)}  a physical part made of the controlled devices equipped with sensors and actuators;
\textit{ii)} a software part made of the (sub-)controllers which interact with the physical part through an event dispatcher.   
This abstraction covers the four-layers architecture widely admitted \cite{DBLP:journals/comsur/Al-FuqahaGMAA15} now for (SOA-based) IoT systems;
the four layers of this architecture are the  \textit{devices, network, services} and \textit{application} layers; going from the physical level (the devices) to the user application level. 



\pgfdeclarelayer{background}
\pgfdeclarelayer{foreground}
\pgfsetlayers{background,main,foreground}

\tikzstyle{sensor}=[draw, fill=green!20, text width=3.8em, 
    text centered, minimum height=2.1em]
\tikzstyle{thing}=[draw, fill=gray!90, text width=3em, 
    text centered, minimum height=1.2em]
\def\blockdist{2.3}
\def\edgedist{2.5}
\tikzstyle{actuator}=[draw, fill=red!60, text width=3.8em, 
  text centered, minimum height=0.4em]
\tikzstyle{service}=[draw, circle, fill=green!80, text width=2.8em, 
    text centered, minimum height=0.2em]
\tikzstyle{arrowRightText} = [above, text width=4em] 
\tikzstyle{dispatch} = [sensor, text width=3em, fill=yellow!60, 
  minimum height=18em, rounded corners]
\tikzstyle{ctrlr} = [draw, text width=5em, fill=blue!40, 
  minimum height=2em, rounded corners]
\tikzstyle{ann} = [above, text width=3em]
\def\blockdist{2.3}
\def\edgedist{2.5}

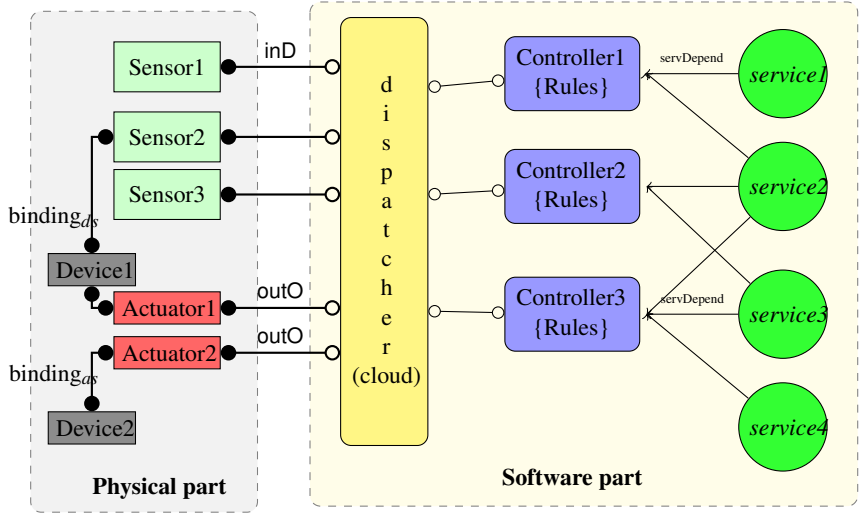
\begin{figure}[!ht]
\begin{tikzpicture}
  \node (dispatcher) [dispatch] {\begin{tabular}{c} d \\ i\\s\\p\\a\\t\\c\\h\\e\\r\\(cloud) \end{tabular}};
  \node (softPart)[below of= dispatcher,yshift=-2.3cm,xshift=2.5cm] {\textbf{Software part}};
    \path (dispatcher.105)+(-\blockdist,0) node (sens1) [sensor] {Sensor1};
    \path (dispatcher.115)+(-\blockdist,0) node (sens2) [sensor] {Sensor2};
    \path (dispatcher.140)+(-\blockdist,0) node (sens3) [sensor] {Sensor3};
    
    \node(dev1)[thing,below of= sens3,xshift=-1.0cm]{Device1};
    \node (ctrDepend)[above of= dev1,xshift=-0.5cm,yshift=-0.3cm] {{\small binding$_{ds}$}};
    \path[draw, *-*,line width=0.2ex] (sens2.west) -| (dev1.north);
    \path [draw, *-o,line width=0.2ex] (sens1) -- node [above] {\textsf{{\small inD}}} 
        (dispatcher.west |- sens1) ;
    \path [draw, *-o,line width=0.2ex] (sens2) -- node [above] {\textsf{ }} 
    (dispatcher.west |- sens2);
     \path [draw, *-o,line width=0.2ex] (sens3) -- node [above] {\textsf{ }} 
    (dispatcher.west |- sens3);
     %
     \path ([yshift=2.1cm, xshift=0.2cm]dispatcher)+(+\blockdist,0)  node (ctrl1) [ctrlr]
           {\begin{tabular}{c}
               Controller1\\
               \{Rules\}
           \end{tabular}};
     \draw [o-o] (dispatcher.73) -- (ctrl1) ;
     \node (serv1) [service, right of = ctrl1,xshift=1.8cm]  {\textit{\small service1}};
       \path [draw,<-] (ctrl1)+(+0.97cm,0) edge node [above]  {{\tiny servDepend}} (serv1);
     
     \path ([yshift=0.6cm, xshift=0.2cm]dispatcher)+(\blockdist,0)  node (ctrl2) [ctrlr]
           {\begin{tabular}{c}
               Controller2\\
               \{Rules\}
           \end{tabular}};
     \draw [o-o] (dispatcher.40) -- (ctrl2) ;
     \node [service,right of=ctrl2,xshift=1.8cm]  (serv2) {\textit{{\small service2}}};
     \path  [draw,<-] (ctrl2)+(+0.99cm,0) -- (serv2);

     \path ([yshift=-1.1cm, xshift=0.2cm]dispatcher)+(\blockdist,0)  node (ctrl3) [ctrlr]
                      {\begin{tabular}{c}
               Controller3\\
               \{Rules\}
                      \end{tabular}};
     \draw [o-o] (dispatcher.-61) -- (ctrl3) ;
     \node [service,right of=ctrl3,xshift=1.8cm]  (serv3) {\textit{{\small service3}}};
     \path  [draw,<-] (ctrl3)+(+0.99cm,0) edge node [above]  {{\tiny servDepend}} (serv3);
     \path  [draw,<-] (ctrl1)+(+0.99cm,0) -- (serv2);
     \path[draw, <-] (ctrl2)+(0.99cm,0) -- (serv3);
      \path[draw, <-] (ctrl3)+(0.99cm,0) -- (serv2);
     
     \node [service,right of=ctrl3,xshift=1.8cm,yshift=-1.5cm]  (serv4) {\textit{service4}};
      \path  [draw,<-] (ctrl3)+(+0.99cm,0) -- (serv4);
      \path (dispatcher.240)+(-\blockdist,0) node (att1) [actuator] {Actuator1};
    \path (dispatcher.250)+(-\blockdist,0) node (att2) [actuator] {Actuator2};
      \node(dev2)[thing,below of= att2,xshift=-1.0cm]{Device2};
      \path[draw, *-*,line width=0.2ex] (att2.west) -| (dev2.north);
          \node (attachDA)[above of= dev2,xshift=-0.5cm,yshift=-0.3cm] {{\small binding$_{as}$}};
    \path[draw, *-*,line width=0.2ex] (att1.west) -| (dev1.south);
      
    \node (physPart) [below of=dev2,xshift=0.9cm, yshift=0.2cm] {\textbf{Physical part}};
    
    \path [draw, *-o,line width=0.2ex] (att1) -- node [above] {\textsf{{\small outO}}} 
        (dispatcher.west |- att1) ;
    \path [draw, *-o,line width=0.2ex] (att2) -- node [above] {\textsf{{\small outO}}} 
        (dispatcher.west |- att2);

    \begin{pgfonlayer}{background}
        \path (sens1.west |- sens1.north)+(-1.1,0.4) node (a) {};
        \path (att2.south -| physPart.east)+(+0.3,-1.9) node (b) {};
        \path[fill=gray!10,rounded corners, draw=black!50, dashed]
            (a) rectangle (b);
        \path (dispatcher.west |-dispatcher.north)+(-0.4,0.2) node (a) {};
        \path (dispatcher.south -| serv4.east)+(+0.4,-0.8) node (b) {};
        \path[fill=yellow!10,rounded corners, draw=black!50, dashed]
        (a) rectangle (b);
    \end{pgfonlayer}

\end{tikzpicture}
\caption{An abstract architectural view}
\label{figure:iot-appli}
\end{figure}

In certain IoT applications, a device called \textit{gateway} or \textit{broker} or \textit{dispatcher} is used to connect the things and their components to a cloud part;
hence data are exchanged through the things and the cloud. 
The gateways are also used for preprocessing and filtering data before the exchanges through the cloud and the involved things.



A \textit{control application} sends orders, signals or alerts to actuators, according to information collected by sensors.
A control application often uses \textit{rules} stated by a human specialist to issue control orders; 
but they can also be based on machine-learning technique, when the rules are computed from a specific database.
Depending on the global states of sensors,  users can send orders to the control application to activate the actuators: this is a feature of manual control application.
In the case of automatic control application, the application itself decides depending on the data collected by the sensors, to activate related actuators or devices.

Therefore the  main components to deal with are:
a physical part made of sensors, actuators, things, a network infrastructure;
a software part made of a control application and potentially specific control or monitoring services.

Additionally, the components interact through low level or application level communication protocols such as WiFi, bluetooth, ZigBee and MQTT\cite{MQTT-OASIS_2015} which is the most popular at application.


\section{Formal Modelling IoT-based Applications}
\label{section:modelling}

\textbf{Notation.} We use set theory and relation notations to structure the component of the models.
Sets are written with capital letters.
A given relation $r$ defined over the sets $S$ and $P$ is written: $r : S \rel P$;
a function $f$ over  $S$ and $Q$ is written: $f : S \fun Q$;
the relational operators {$ran$} and {$dom$} denotes respectively the \textit{range} and the \textit{domain} of a relation or function. the notation $\power(S)$ denotes the powerset of $S$.\\

An IoT-based application  is composed at least of:
a set of connected IoT devices (${\mathcal D}$), sensors  (${\mathcal S}$) and actuators  (${\mathcal A}$), that makes the physical part;
a set of controllers  (${\mathcal C}$), sometimes together with a dedicated server (or dispatcher) which collects the data from sensors and distributes them to the controllers.
A controller is linked to the sensors from which it reads inputs, and to the actuators which it manages.
The MQTT protocol is often used for the communication between sensors and controllers on the one hand and between the controllers and the actuators on the other hand.

In the case of a simple control application, sensors are connected to a controller which is connected to the actuators.
In the general case of control applications, sensors and actuators are connected to several controllers which are served by a dispatcher which collects the inputs from the sensors and dispatch them to the involved controllers. 

\subsection{The Basic Components of the Model}
\paragraph{Sensors.}
A sensor $s$ is a device that provides a value in a given range; these values correspond to a physical sensing of the environment of the sensor. More specifically a sensor is dedicated to a given device or environment (for instance a room, a light, etc).
The range of values are associated to specific sensors and for diverse purposes; for instance we have temperature sensors, motion sensors, light sensors, contact sensors, etc.
A given value measured by a sensor will correspond to a state of the device that it senses; for instance depending on the value measured by the dedicated sensor a light state will be on or off. 
Each category ($c_s$) of sensors may have various value ranges ($R_{cs}$).
A sensor may interact with its environment (devices, controllers) through one or several communication protocols. Let $\mathit{CommProto}$ be such a set of communication protocols. 

Accordingly a sensor $s$ of the set of sensors ${\mathcal S}$ ($s \in {\mathcal S}$) is defined by a 4-tuple $(c_s,r_{sc},v_s,\mathit{comm_{ps}})$ with 
a category $c_s \in C_s$,
a range $r_{sc} \subseteq R_{cs}$,
a value  in its range $v_s \in r_{sc}$ and
a set of communication protocols $\mathit{{comm_p}_s}$.
We will use the following functions to get each element of the 4-tuple defining a sensor:
\begin{center}
  \begin{tabular}{rcl}
$\mathit{categ}_s: {\mathcal S} \fun C_S$ ~& $~~~~$ & $\mathit{range}_s : {\mathcal S} \rel R_{cs}$\\
  $\mathit{value}_s: {\mathcal S} \fun R_{cs}$ ~& & $ \mathit{comm_p}_s :  {\mathcal S} \rel {CommProto}$
\end{tabular}
\end{center}

We will describe later the links between a sensor and a given device or a controller.

\paragraph{Actuators.}
An actuator $a$  of the set of actuators ${\mathcal A}$ ($a \in {\mathcal A}$) is a device that receives an order from a human or a controller, and sends accordingly a signal to its environment, which can be a physical object or a device. For instance, upon the reception of an order \textsf{on}, a light actuator may send an impulsion signal to put the light on.
An actuator $a$ receives an input in a specific range of order values ($Order_A$), and provides accordingly an output signal ($Signal_A$) towards its environment.

We use the following relations to determine the elements of the triple $(i_a, o_a, p_a)$ that describes an actuator a:
$$\mathit{inputOrd}_a: {\mathcal A} \rel Order_A ~~~~
\mathit{ouputSign}_a: {\mathcal A} \rel Signal_A ~~~~
\mathit{comm_p}_a :  {\mathcal A} \rel CommProto$$
They give the set of inputs, outputs and protocols of an actuator. In the same way as for sensors, we will define later the links between an actuator and a device or a controller.

\paragraph{Devices.}
A device ($d \in {\mathcal D}$) is modelled at any time by its state $s_d$ in such a way that, a range of measured values $V_v$ of some sensors $s$  (which are linked to the device $d$), corresponds to this state $s_d$ (that is $V_v \mapsto s_d$); for example, a light state will be set to \textsf{on} according to the sensed dimmed values between a given range.
In the same way the output signals of an actuator can, upon the reception of an order from a controller, set the device $d$ in a state $s_d$; for example a controller can put \textsf{off} the light, resulting in the actuator setting the power (electrical energy) of the light to 0.

As a consequence the device $d$ is characterised by its set of states (${\mathit STATED}$), corresponding to its behaviour which is moving from state to state according to the received stimulation signals; that is a labelled transition system $\langle {\mathit STATED},~ S_0,~ {\mathit SignalA},~ \delta \rangle$, with ${\mathit SignalA}$ the set of received signals and $\delta :{\mathit STATED} \times {\mathit SignalA} \fun {\mathit STATED}$, and $S_0$ an initial state of the device. 
Accordingly a device $d$, without the links with its environment which will be defined later, is modelled by its set of states, a current state ($curState_d$)  which is initially $S_0$, and a transition system which abstracts its behaviour.
\begin{center}
  \begin{tabular}{c}
    $curState_d :  {\mathcal D} \fun STATED$\\
    $behaviour_d :  {\mathcal D} \fun \langle {\mathit STATED},~ S_0,~ {\mathit SignalA},~ \delta \rangle$
  \end{tabular}
  \end{center}


Here is an illustration: when we turn a dimmer switch (actuator), the lighting of a dimmed light (device), will going more and more bright or dark; a light sensor connected to the light may show weaker or higher measured lumens that correspond to the effective lighting intensity. But the state of the light will be either \textsf{on} or \textsf{off}. 
\paragraph{Control and communications.}
The tasks performed in controller applications consist in a set of rules ($R$) that are applied to analyse data ($D_S$) collected from the sensors (${\mathcal S}$) and to synthesise accordingly the orders (${\mathcal Order}$) to be sent to the actuators.
A controller $c$ is then equipped with a function ${\mathit ComputeOrder_R}$ defined on $D_S  \fun  {\mathcal Order}$; when the controller collects values from sensors binded to it, it computes the appropriate order, 
and outputs  this order to the actuators binded to it (to act on their environment).
For instance a controller $c$ sends an order \textsf{on} to a light actuator $a$ when the related sensor detects darkness if $c$ has a rule which states that the light should be put on in case of darkness.
Sensors and actuators may be grouped to form a (wireless) sensor-actuator network;
accordingly, the controllers interact with a gateway to which the sensors and actuators are linked via their network.

\subsection{Abstract Model of IoT-based Applications}
\label{section:abstractModel}

We describe an abstract model of an IoT-based application according to the two main components presented in Section \ref{section:background}; we build the abstract model $M_{phys}$ of its physical part and  the abstract model
$M_{soft}$ of its software control part.
For the physical part, we consider
${\mathcal S}$ as a set of sensors,
${\mathcal A}$ as a set of actuators,
${\mathcal D}$ as a set of devices.

The physical architecture of an IoT system is modelled with a n-tuple $$M_{phys} = (S,~ A,~ D,~  binding_{(d,s)},~ binding_{(a,d)})$$
where $S \subseteq {\mathcal S}$ is a subset of sensors;
$A \subseteq {\mathcal A}$ is a subset of actuators;
$D \subseteq {\mathcal D} $ is a set of sensed or controlled devices;
$binding_{(a,d)} \subseteq  A \times D$ is a relation that  describes the binding between the actuators and the controlled devices and
$binding_{(d,s)} \subseteq  D \times  S$ is a relation that  describes the binding between the sensed devices and their sensors.

A control software part (or control application) of an IoT system is modelled with the tuple $$M_{soft} = ({\mathcal C_R}, Serv, servDepend_{(c,s)})$$
where
${\mathcal C_R} \subseteq {\mathcal C}$ is a set of controllers which use a set of the control rules $R$ for their control tasks; 
$Serv$ is a set of services used or provided by the controllers of the control application;
$servDepend_{(c,s)} \subseteq {\mathcal C_R} \times Serv$ is the dependencies between  the controllers and the used services. 

To model the link between the control application and the sensors and actuators of the physical IoT architecture, the physical and software models ($M_{phys}, M_{soft}$) are linked with the following relations: 
\begin{itemize}
\item $inD \subseteq S \times C$  which models the link between the sensors and the controller; it supports data input from the sensors;
\item $outO \subseteq C \times A$  which models the link between the controller and the actuators; it supports order output to actuators.
  \end{itemize}

At this stage the control application ($M_{soft}$) can communicate with the physical architecture via the abstract model ($M_{phys}$) of this architecture. 
Typically the application receives data from the sensors (via  $inD$) and issues orders sent to the actuators (via $outO$).
The orders are computed from the application services using the defined rules.

But, in order to state the architectural invariants and to analyse the IoT system properly, we should fix one of the identified shortcomings leading to inconsistencies, which is the lack of explicit declaration of dependencies between sensors and controlled devices.  
When the control of a given device depends on some sensors, this dependency should be made explicit.
The devices should have been equipped by an actuator which share the same controller with the involved sensors.
Therefore we require to make explicit in the model,  the control dependency relation between involved sensors and controlled devices with a relation $CtrlDepend_{(s,d)} \subseteq S \times D$.

This relation describes which sensors impact which devices, so that we can reason later on the consistency of the functioning of the global system.
The relations $CtrlDepend_{(s,d)}$  and $binding_{(d,s)}$ should not be confused since the impacted devices described by $CtrlDepend_{(s,d)}$ can be different from the sensed ones described by $binding_{(d,s)}$.    

\noindent
Consequently,
given a physical architecture $M_{phys} = (S,~ A,~ D,~  binding_{(d,s)},~ binding_{(a,d)})$,
a control part $M_{soft} =({\mathcal C_R},~ Serv,~ servDepend_{(c,s)})$
and their interconnection  with the relations  $inD$, $outD$ and  $CtrlDepend_{(s,d)}$,
the global model of the complete control system $Sys$ integrating the parts $M_{phys}$ and $M_{soft}$ is described by the 5-tuple: 
$$Sys = (M_{phys},~ M_{soft},~ inD,~ outD,~ CtrlDepend_{(s,d)})$$
More importantly, for practical reasons i.e. the systematic construction and the analysis of the model, we propose to define all or some parts of the global system as parameters:\\
$Sys[M_{phys},~ M_{soft},~ inD,~ outD,~ CtrlDepend_{(s,d)}]$.
This enables one to build separately the different parts, and also to modify them easily as well as their interconnections; for instance we can fix a physical architecture and check different versions of the control part or as done in the following, fix the software and check some configurations of the physical parts.

Hence, we denote the global model by a parametrised structure raising $M_{phys}$, $inD$, $outD$ and  $CtrlDepend_{(s,d)}$ as the parameters of the model, and fixing a software part: 
$$Sys_{(M_{soft})}[M_{phys},~ inD,~ outD,~ CtrlDepend_{(s,d)}]$$
Note that from this stage, a specific concrete domain specific language (DSL) can be built from our abstract modelling; we introduce such a DSL in subsection \ref{section:practice}.
Reciprocally a mapping can be made between the abstract model and existing IoT DSL.
\subsection{Behavioural Description of a Control Application}
A control application continuously reacts to the data collected by sensors and change the state of its environment  by sending orders to the involved actuators which act on the thing or the environment.
When there is no collected data or no specific order to change the state of the system, the control application stays passive.

We use operational semantics rules to describe the behaviour of the control applications.
First, we assume that the application is \textit{consistent} so that, it can react properly to the sensed data.
In the next section (Sect. \ref{section:consistencyProp}) we show how the consistency properties are defined and then how it can be checked (Sect. \ref{subsection:consistencyAnalysis}). 

Given a consistent  application  $Sys_{(M_{soft})}[M_{phys},~ inD,~ outD,~ CtrlDepend_{(s,d)}]$,
with $M_{phys} = (S,~ A,~ D,~  binding_{(d,s)},~ binding_{(a,d)})$ and
$M_{soft} =({\mathcal C_R},~ Serv,~ servDepend_{(c,s)})$, 
when a controller $c_i$ (with a function ${\mathit computeOrder}_{c_{i}}$) of $C_R$,  receives a value \textsf{val} from a sensor $s_i$ of $M_{phys}$ binded to a device $d_s$ of $M_{phys}$, considering that the control of a device $d_c$ depends on the sensor $s_i$, and that there is an actuator $a_s$ binded to $d_c$, then an order \textsf{ord}$_i$, computed by the controller $c_i$  linked to $a_s$, is sent to the actuator $a_s$.

Consequently we formally define the general behaviour of the application by the following operational semantic rule.
It captures well the traditional \textsf{sense-decision-control} paradigm of control systems.

$${\frac{
    \begin{tabular}{c}
      $M_{soft} =({\mathcal C_R},~ Serv,~ servDepend_{(c,s)})$\\
      $M_{phys} = (S, A,  D,  binding_{(d,s)}, binding_{(a,d)})$\\
    $Sys_{(M_{soft})}[M_{phys},~ inD,~ outD,~ CtrlDepend_{(s,d)}]$ \\
      $c_i \in C_R ~~~~ s_i \in S ~~~~a_s \in A~~~~d_s \in D~~~~d_c \in D$ \\
      $(d_s, s_i) \in binding_{(d,s)}~~~~(a_s, d_c) \in binding_{(a,d)}$ \\
      $(s_i, c_i) \in inD ~~~~~ (c_i, as) \in outO ~~~~~ (s_i,d_c) \in CtrlDepend_{(s,d)}$\\
      $c_i\Downarrow(s_i, \textsf{val})   ~~~~  \textsf{val} \in range_s(s_i) ~~~~~$
      $\textsf{ord}_i = {\mathit ComputeOrder}_{c_{i}}(\textsf{val})$
    \end{tabular}
}{
      c_i\Uparrow(a_s, \textsf{ord}_i)} }{(\sf reactOnSense)}
$$

The operators $\Downarrow$ and $\Uparrow$ denote respectively
the reception of a value from a given sensor by a controller  and
the sending of an order by a controller to an actuator.
Thus $c_i\Downarrow(s_i,  \textsf{val})$ expresses that the controller $c_i$ receives the value \textsf{val} sent by the sensor $s_i$; similarly $c_i\Uparrow(a_s, \textsf{ord}_i)$ expresses that the controller $c_i$ sends the order \textsf{ord}$_i$ to the actuator $a_s$.

A consequence of the previous rule is the \textit{Integrity of orders}:
any order sent to an actuator results from one of the services of the control application.
As the computations of orders are due to the controller whose services implement the control application rules ($R$), the orders sent to the actuators should be the right ones. However the integrity checking may be deeply propagated till the application implementation level.   


\section{Consistency Properties and Analysis of the Formal Model}
\label{section:consistencyModel}
Here, we enhance the generic formal model built in the previous section with required consistency properties.

\subsection{Invariant Consistency Properties}
\label{section:consistencyProp}
We focus in this section on consistency correctness concerns;
that is the architectural consistency and then the consistency of the functioning of IoT-based applications.

Each of the following rules expresses a property that we can deduce from a consistent model.
Reciprocally a model satisfying these properties will be consistent.

\paragraph{Connection of physical components.} Any consistent IoT architecture has connected devices, sensors and actuators.

$$
{\frac{\begin{tabular}{c}
          $Sys_{(M_{soft})}[M_{phys},~ inD,~ outD,~ CtrlDepend_{(s,d)}]$ \\
      $M_{phys} = (S, A,  D,  binding_{(d,s)}, binding_{(a,d)})$\\
    \end{tabular}
  }
  {  binding_{(d,s)} \neq \emptyset ~~\land~~ binding_{(a,d)} \ne \emptyset}
}
{\sf (connectedHWCpnts)}
$$

\paragraph{Well-structuring of controllers.} Any consistent control application is linked to at least one sensor and one actuator.
$$
{\frac{\begin{tabular}{c}
    $Sys_{(M_{soft})}[M_{phys},~ inD,~ outD,~ CtrlDepend_{(s,d)}]$ \\
      $M_{soft} =({\mathcal C_R},~ Serv,~ servDepend_{(c,s)})$\\
      $c_i \in C_R$
    \end{tabular}
  }
  {inD(c_i) \neq \emptyset ~~\land~~outO(c_i) \ne \emptyset}
}
{\sf (FPwellStructCtrl)}$$

\medskip
\paragraph{Consistency of components involved in interactions.}
Sensors or actuators involved in the interactions are those described in the physical support.

$$
{\frac{\begin{tabular}{c}
    $Sys_{(M_{soft})}[M_{phys},~ inD,~ outD,~ CtrlDepend_{(s,d)}]$ \\
    $M_{phys} = (S, A,  D,  binding_{(d,s)}, binding_{(a,d)})$
    \end{tabular}
  }
  {\dom(inD) \subseteq S ~~\land~~\ran(outO) \subseteq A}
}
{\sf (FPweakConsistentCpnts)}$$

 But, this consistency is weak, because it does not constrain the linking of the involved sensors and actuators.

To be more accurate, we need a more strong property which should state that:
the actuators to whom a controller sends its orders (via $outO$),
are those actuators binded (via $binding_{(a,d)}$) to the devices  which are controlled  (via $CtrlDepend_{(s,d)}$) by the sensor binded (via $inD$) to the controller. 
Thus the interaction consistency property is established through the following commutative diagram; it states an invariant property:\\

\begin{multicols}{2}
\begin{tikzpicture}{(\textsf{FPconsistBindings})}
  \matrix (m) [matrix of math nodes,row sep=3em,column sep=5.5em,minimum width=2em]
  {
     \mathit{SENSOR} & \mathit{DEVICE} \\
     \mathit{CONTROLLER} & \mathit{ACTUATOR} \\};
  \path[-stealth]
    (m-1-1) edge node [left] {$inD$} (m-2-1)
  edge node [below] {$CtrlDepend_{(s,d)}$} (m-1-2)
      (m-2-1.east|-m-2-2) edge node [below] {$outD$} (m-2-2)
    (m-1-2) edge node [right] {$binding_{(a,d)}^{-1}$} (m-2-2);
\end{tikzpicture} 

$~~$\\
\vspace{2.5cm}
\hspace{2cm} (\textsf{FPconsistBindings})
\end{multicols}

A consequence of the previous property is the \textit{Consistency of control dependencies.}
If a sensor $s_i$ impacts the control of a given device $d_i$ (via $CtrlDepend_{(s,d)}$),
and the sensor $s_i$ is connected to a controller $c_i$ (via $inD$) 
then the actuator $a_k$ binded (via $binding_{(a,d)}$) to the device $d_i$ is also linked (via $outO$) to the controller $c_i$:
$$
{\frac
  { \begin{tabular}{c}
      $M_{soft} =({\mathcal C_R},~ Serv,~ servDepend_{(c,s)})$\\
      $M_{phys} = (S,~ A,~ D,~  binding_{(d,s)},~ binding_{(a,d)})$\\
      $Sys_{(M_{soft})}[M_{phys},~ inD,~ outD,~ CtrlDepend_{(s,d)}]$ \\
      $s_i \in S ~~~~~~ c_i \in C_R ~~~~~~ a_k \in A  ~~~~~~ d_i \in D$\\
      $(s_i, d_i) \in CtrlDepend_{(s,d)}  ~~~~~ ~~~ (s_i, c_i) \in inD ~~~~(a_k,d_i) \in binding_{(a,d)}$\\
  \end{tabular}}
  { (c_i,a_k) \in outO}
}
{(\sf FPCtrlDependency)}
$$

\medskip
\begin{proof}
  From the commutative diagram, we infer that $\forall s_i \in S$,\\
  if $\exists d_s \in D,~ \exists a_k \in A | (s_i, c_i) \in CtrDepend_{(s,d)}~ \land~ (d_s, a_k) \in binding_{(a,d)}^{-1}(d_s,a_k)$ \\
  then $\exists c_u \in C_R,~ \exists a_w \in A | (s_i, c_u) \in inD ~\land~ (c_u, a_w) \in outO$\\
  the diagram commuting, $c_u$ and $a_w$ are unique, hence $c_u = c_i$ and $a_w = a_k$\\
  and we have $(c_i, a_k) \in outD$
  \qed
\end{proof}

\medskip
\paragraph{Well-connection of actuators and sensors.}  
The controllers which are connected to sensors should also be connected to some actuators, otherwise the collected inputs are not used for the control:
$$
{\frac{\begin{tabular}{c}
      $Sys_{(M_{soft})}[M_{phys},~ inD,~ outD,~ CtrlDepend_{(s,d)}]$ \\
      $M_{phys} = (S,~ A,~ D,~  binding_{(d,s)},~ binding_{(a,d)})$\\
    \end{tabular}}
  { \ran(inD) \subseteq \dom(outO) }
}
{(\sf FPsensor2Actuator)}
$$

\medskip
\paragraph{Communication protocols.}
The pairs of sensor-controller and controller-actuator use compatible communication protocols:
each sensor interacts with the binded controller using an appropriate communication protocol;
each  controller interacts with the binded actuators using an appropriate communication protocol.
Here we consider the set of communication protocols used by the components of the architecture, and we manage compatibility with sets inclusion.

$$
{\frac{ \begin{tabular}{c}
      $M_{soft} =({\mathcal C_R},~ Serv,~ servDepend_{(c,s)})$\\
      $M_{phys} = (S,~ A,~ D,~  binding_{(d,s)},~ binding_{(a,d)})$\\
      $Sys_{(M_{soft})}[M_{phys},~ inD,~ outD,~ CtrlDepend_{(s,d)}]$ \\
      $c_i \in C_R ~~~ s_i \in S ~~~a_s \in A~~ (s_i, c_i) \in inD~~~~(c_i,a_i) \in outD$\\
      ${comm_p}_s(s_i) \subseteq \mathit{CommProto} ~\land~ {comm_p}_c(c_i) \subseteq \mathit{CommProto}~\land~$\\
      ${comm_p}_a(a_s) \subseteq \mathit{CommProto}$
      \end{tabular}
  }
    { {comm_p}_s(s_i)~ \cap~ {comm_p}_c(c_i) \neq \emptyset ~~~~\land~~~~
       {comm_p}_c(c_i)~ \cap~ {comm_p}_s(a_i) \neq \emptyset }
}
{(\sf FPCompComm)}
$$

\subsection{Consistency Analysis of IoT-based Control Applications}
\label{subsection:consistencyAnalysis}
Given the previous defined consistency rules, we state the following propositions for the analysis of IoT-based control applications. The idea is the backward exploitation of the rules: if a given model satisfies the rules then it is consistent. 

\begin{proposition}{(Architectural correctness)}
  A given architecture $M_{phys}$ is consistent if the property
 {\rm \textsf{connectedHWCpnts}}    is satisfied. 
\end{proposition}

\begin{proposition}{(Correctness of functioning)}
  An application parametrised with $M_{phys}$, $inD$, $outO$ and $CtrlDepend_{(s,d)}$ is consistent if:
  $M_{phys}$ is consistent and
  the properties {\rm
    \textsf{FPwellStructCtrl},
    \textsf{FPsensor2Actuator},
    \textsf{FPCompComm},
    \textsf{FPweakConsistentCpnts},
    \textsf{FPconsistentBindings},
    \textsf{FPCtrlDependency} }
 are satisfied. 
\end{proposition}

Consequently if we build a model having these properties as invariants, then the model is consistent by construction. This is the basic idea in the following section.

\section{Checking the Consistency Properties using Event-B}
\label{section:checkingArchitecture}

We propose a generic formal framework to implement and analyse the models of  given architectural descriptions of IoT-based systems. The Event-B formalism and method are used for this purpose. 

\subsection{Overview of Event-B Models Structuring}

Event-B models are structured with machines and refinements.
An event-B machine has a context, a state space description using variables and invariants and a list of events.
Several machines may share the same contexts, therefore the context is often defined as a standalone structure comprising sets, constants, axioms. A context is the seen by a machine. Like a machine, a context can be extended to build a larger context.
A refinement is a more concrete machine that refine an abstract machine. A refinement can see a context.

Event-B \cite{EventB_Abrial2010,DBLP:journals/scp/HoangKBA09} is a modelling and development method where components are modelled as abstract machines which are composed and refined into concrete machines. 
An \textit{abstract machine}  describes a mathematical model of a system behaviour\footnote{A system behaviour is a discrete transition system}.
In an Event-B modelling process, abstract machines constitute the dynamic part whereas \textit{contexts} are used to describe the static part.  
 A \textit{context} is seen by machines. It is made of carrier sets and constants.
 It may contain properties (defined on the sets and constants), axioms and theorems.
A machine is described, using properly named clauses, by a state space made of typed variables and invariants, together with several \textit{event} descriptions. 

\vspace{-0.2cm}
\paragraph{State Space of a Machine}
The variables constrained by the invariants (typing predicates, properties) describe the  state space of a machine.
The transition from one state to the other is due to the effect of the events of the machine. Specific properties required by the model may be included in the invariant. The predicate $I(x)$ denotes the invariant of machine, with $x$ the list of state variables.

\vspace{-0.2cm}
 \paragraph{Events of an Abstract Machine}
Within Event-B, an event is the description of a system transition. Events are  spontaneous and show the way a system evolves. 
An event $e$ is modelled as a \textit{guarded substitution}: $e \defs eG \Longrightarrow eB$ where $eG$ is the event \textit{guard} and $eB$ is the event \textit{body} or \textit{action}.
An event may occur only when its guard holds. 
The action of an event describes, with simultaneous generalised substitutions, how the system state evolves when this event occurs: disjoint state variables are updated simultaneously.

The effect of events are modelled with generalised logical substitution (S) using the global variables and constants. For instance a basic substitution  \texttt{x := e}  is logically equivalent to the predicate \textit{x' such that x' = e}. This is symbolically written $ x' : (x' = e)$ where $x'$ corresponds to the state variable $x$ after the substitution and $e$ is an expression. In the rest of the paper, the variable $x$ is generalised to the list of state variables.

Several events may have their guards held simultaneously; in this case, only one of them  occurs. The system makes internally a nondeterministic choice. If no guard is true the abstract system is blocking (deadlock).

In Event-B \textit{proof obligations} are defined to establish model consistency via invariant preservation.
Specific properties (included in the invariant) of a system are also proved in the same way.

\vspace{-0.2cm}
\paragraph{Refinement.} An important feature of the Event-B method is the availability of refinement technique to design a concrete system from its abstract model by stepwise enrichment of the abstract model. During the refinement process new variables ($y$) are introduced; the invariant is strengthened without breaking the abstract invariant, and finally the events guards are strengthened. In the invariant $J(x,y)$ of the refinement, abstract variables ($x$) and concrete variables ($y$) are linked. The refinement is accompanied with proof obligations in order to prove their correctness with respect to the abstract model.

\vspace{-0.2cm}
\paragraph{\texttt{Rodin} Tool.} \texttt{Rodin}\footnote{{\small http://wiki.event-b.org/index.php/Main\_Page}}  is an open tool dedicated to building and reasoning on B models, using mainly provers and the \textsf{ProB} model-checker. \texttt{Rodin} is made of several modules (plug-ins) to work with B models and interact with related tools.\\

\vspace{-0.3cm}
\subsection{A Generic Framework for Consistency Checking}
As described in Sect. \ref{section:abstractModel}, the common requirements and properties of IoT-based applications are captured through an abstract generic model;  the analysis of the consistency properties does not depend on a specific application and can be done through a generic framework.\\ 
\vspace{-0.6cm}
\begin{figure}[ht]
  \begin{center}
      \includegraphics[height=0.4\textheight,width=0.7\linewidth]{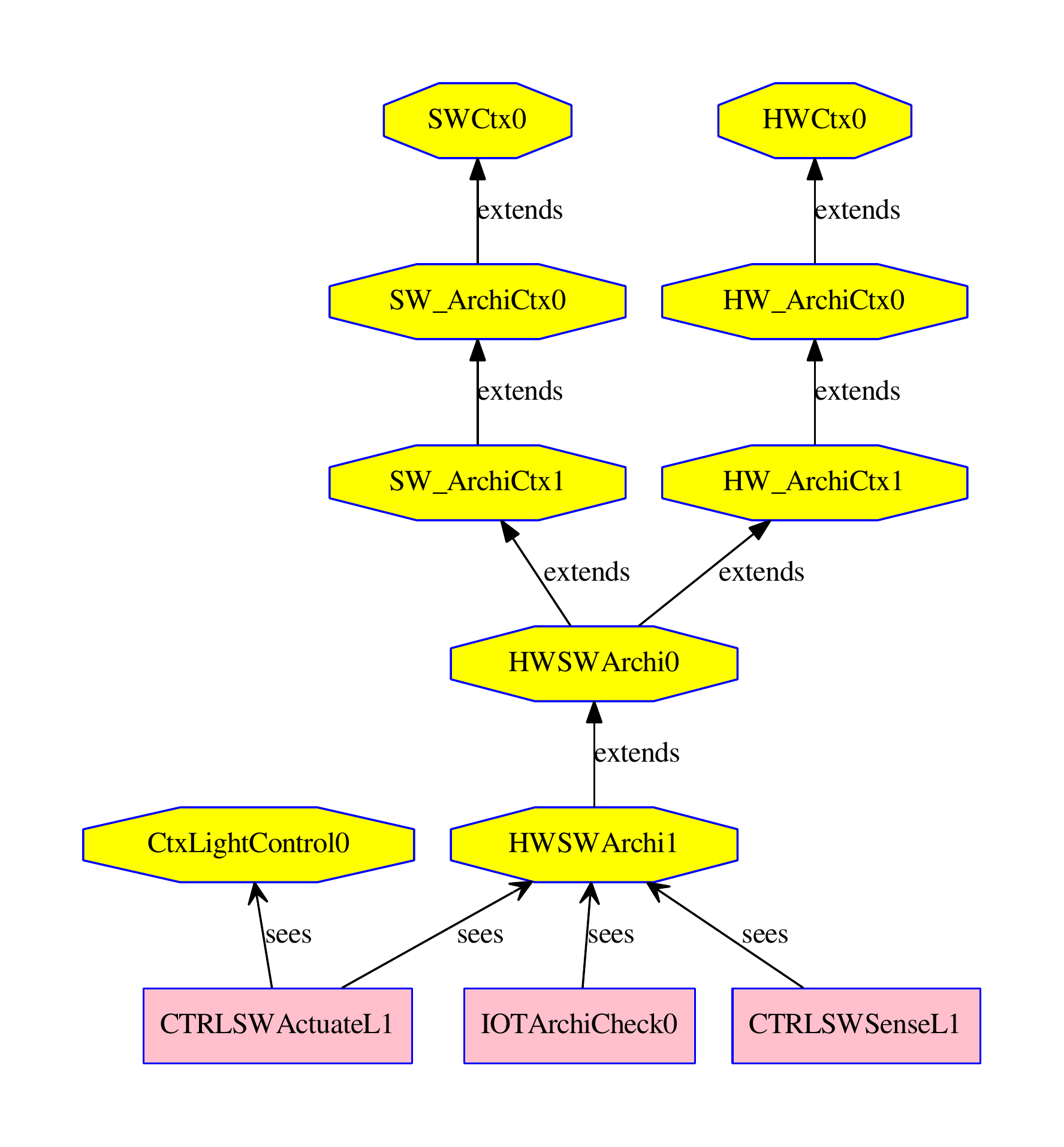}  
      \caption{A generic Event-B modelling and analysis architecture}
  \label{figure:eventB_iot_archi}
  \end{center}
\end{figure}

\vspace{-0.6cm}
To gain extensibility, we consider families of IOT-based applications; for instance a home control family where the main components of applications are always the same: lights, windows, doors, garage, heating, etc. Therefore the modelling components are not varying and can be gathered as reusable components in our formal modelling. That explains the structuring of our generic architecture where some contexts and machines are to be adapted to specific applications but other machines are defined once for all.

We implement the generic analysis framework in Event-B (see Fig. \ref{figure:eventB_iot_archi}\footnote{We have implemented a tool that generates this diagram from any Event-B project}) following the structure of the abstract model that is a parametrised structure interconnecting in a systematic way, a physical part and a control software part. 
The framework is not only designed and used to implement our proposed method of modelling and analysis, it aims at being an easily reusable framework.
For this purpose, we adopted a layered structuring of the framework in order to have a systematic approach for building, generating or extending the framework. 
A basic layer comprises fixed predefined Event-B contexts (octagons whose names end with 0) which gather all elementary types and relations required in any application withing a given family. Another layer comprises Event-B contexts and machines (whose names end with 1) which are the specific instantiations of the predefined context.  
Considering for instance a family of home automation applications,
at the hardware level, the context \textsf{HWCtx0}  contains  the basic sets (LIGHTSENSOR, MOTIONSENSOR, LIGHTACTUATOR, etc) for the family; 
at the software level the context \textsf{SWCtx0} contains all the basic sets (SERVICE, CONTROLLER) for the applications of this family.

The context \textsf{HW\_ArchiCtx0} implements $M_{phys}$; it contains the generic structuring of a physical architecture (the formal bindings between the devices; it contains the relations $binding_{(d,s)}$ and $binding_{(a,d)}$ (see Sect. \ref{section:abstractModel});
Similarly \textsf{SW\_ArchiCtx0} implements $M_{soft}$; it contains the generic structuring of the software part (with $servDepend$).
The context \textsf{HW\_ArchiCtx1} contains a specific instantiation for the physical architecture.
It comprises the declarations of the objects (the sensors of each type, the needed controllers, etc) and their assembly in a given application.
Similarly \textsf{SW\_ArchiCtx1}  contains a specific instantiation  of the software part; it comprises the controllers and the services on which they depend.
Thus, only these two contexts will be modified to consider new instances of physical or software part.
The generic interconnection between the two parts (with the relations $inD$, $outO$, $CtrlDepend$, see Sect. \ref{section:abstractModel}) is implemented  with the Event-B context \textsf{HWSW\_Archi0}.
In the same way, the context \textsf{HWSW\_Archi1} models a specific instantiation; it is the only context to be modified in order to build an interconnection for a specific control application; it gathers  $M_{soft}$ and the four parameters ($M_{phys}$, $inD$, $outO$, $CtrlDepend$). The  \textsf{CTRLSWActuateL1} machine is thus one example of system parametrised by \textsf{HWSW\_Archi1}. 

The analysis machine (\textsf{IoTArchiCheck0}), defined once for all, contains the invariant properties to be checked for any given interconnection of physical and software part, which is given as a parameter (the \textsf{HWSW\_Archi1} machine). That explains the structuring with the Event-B \textsc{sees} clause: a way to implement the genericity. Note that the development of an example system, is orthogonal to the checking.
The machine \textsf{IoTArchiCheck0} contains all the properties to be checked:
\textsf{FPwellStructCtrl},
\textsf{FPweakConsistentCpnts},
\textsf{FPconsistBinding},
\textsf{FPCtrlDependency},
\textsf{FPsens2actu}
(see Sect. \ref{section:consistencyProp}).
If the model in the \textsf{IoTArchiCheck0} machine is proved correct, then all the architectural and consistency properties are satisfied and consequently the model is consistent.
We use the \textsf{Rodin} to perform  the proofs; the aim is, given a model describing an IoT-based application, to prove at least all the properties of interest which are listed above. 

We experiment the framework with some examples, as shown in the following.
\vspace{-0.3cm}
\subsection{Putting into Practice and Assessment}
\label{section:practice}
To detail for what use and how to use the proposed framework, 
let us consider a description of an IoT-based application, including an interconnection between a physical part and a software part.  First, the framework enables ones to capture the formal model of this application. Second, it enables ones to check the captured model with respect to the invariant properties we have identified.
For this purpose, we have to describe the application at hand as a parameter of our generic framework. That is:
\textit{i)} to set the physical description in the \textsf{HW\_ArchiCtx1} context,
\textit{ii)} to set the software part description in the \textsf{SW\_ArchiCtx1} context;
\textit{iii)} to set the desired interconnection in the \textsf{HWSWArchi1} context. 
After setting these data in the Event-B model, checking the \textsf{IOTArchiCheck0} results in a success if the described application is consistent, otherwise it is not. We have automatised the full process.

We apply the method with several scenario and it works well. Notably the \textsf{FPconsistentBindings} property appears to be very helpful. The illustrative example below reveals a failure due to the inconsistency of the required control dependency between the light sensor \texttt{ls2} and the light \texttt{la}.

\vspace{-0.2cm}
\paragraph{An Illustrative example description.}
To facilitate the use of the framework, we design a tiny Domain Specific Language, from which we generate the Event-B contexts \textsf{HW\_ArchiCtx1}, \textsf{SW\_ArchiCtx1}, \textsf{HWSWArchi1} to be used for a given analysis experimentation. It is as follows.
\begin{multicols}{2}
  \noindent
  \begin{boxedminipage}{6.0cm}
    {\small
  \begin{alltt}
IOTSystem  ExampleApp 
// Physical part 
LIGHTSENSOR   : ls1, ls2
LIGHTACTUATOR : la
LIGHT         : lvrl1

// HW Architecture
ADBinding : (lvrl1, ls2)
DSBinding : (la,  lvrl1)
\end{alltt}
    }
  \end{boxedminipage}
  
\noindent
  \begin{boxedminipage}{6.0cm}
    {\small
  \begin{alltt}
//PhysSoftInterconnection
SCBinding: (ls1, ctl1), (ls2, ctl2)
CABinding: (ctl1 , la)
SDDependeny : (ls2, lvrl1)
\end{alltt}
}
\end{boxedminipage}

 \noindent
 \begin{boxedminipage}{5.8cm}
    {\small
  \begin{alltt}
// Control part 
CONTROLLER : ctl1, ctl2
SERVICE : srv1, srv2
// SW Architecture
Control-Service : (ctl1, srv1)

// Behavioural Rules
srv1 : \{
Lightvalue(n) --> Order(on) 
Lightvalue(0) --> Order(off)
Lightvalue(m) --> Order(on) 
\}
srv2 : \{
DoorValue(...)  --> Order(...)
DoorValue(...)  --> Order(...)
\}
\end{alltt}}
\end{boxedminipage}  
\end{multicols}

\vspace{-0.2cm}
\paragraph{Analysis results.}
Using our method, we quickly detect architectural inconsistencies which will lead to a dysfunction of a control system, due to a sneak inconsistency; indeed sensors and actuators can be working perfectly but the control system can send right orders to the wrong actuators.
The analysis of the abstract model raises such an anomaly at design time.
\vspace{-0.4cm}
\subsection{Related Work}
There are several works dedicated to the modelling and the analysis of IoT applications as we have done; they often take into account a specific concern, and as such can be considered as complementary; but in our knowledge there is no widely shared abstract model that can help the interoperability between the existing proposals and results. We target this objective by proposing, compared to some of the existing work, an open and extensible abstract model. 
In \cite{design_analysis_iot_2016} the authors introduce SysML4IoT to define a model compliant with the IOT-A reference model, and they translate the SysML model into NuSMV programs for the analysis concern. Their focus was on the verification of quality of service (QoS) properties.
The authors of \cite{mbaVerifFGCS2019} focus  on a multiview modelling together with workflows for implementing cloud-based Industrial IoT systems. For the modelling they combine several views through various models, and integrate them using the Automation Markup Language; they chose Uppaal for verification aspects and combine the  Uppaal Timed Automata models with action patterns of timing behaviour  to verify the timing performance to guarantee timing properties.
%
The concerns of \cite{Fattah2017BuildingIS,DBLP:journals/computer/SalahuddinAGSS17} are related to IoT services for health-care.
In \cite{enablingHL_IOT_2015} the authors propose a development methodology and an associated framework to ease the development of IoT applications, but formal analysis was not their concern.
%
In \cite{SpecVerifMQTT2017,FormalModelMQTT_IEEE2014} verification of communication protocols such as MQTT are dealt with. Timed process-algebra \cite{FormalModelMQTT_IEEE2014} and Probabilistic timed automata and statistical model checking in \cite{SpecVerifMQTT2017} uses are used for this purpose.

\section{Conclusion}
\label{section:conclusion}
We have designed a generic formal model together with architectural and consistency properties that have been formalised as the invariants that characterise many IoT-based applications.
We then proposed  a generic framework for  the formal modelling of IoT-based applications, and the rigorous analysis of their consistency properties.
The framework is structured in a generic way by distinguishing different parts which serve as parameters in order to favour extension and reusability. 
We have shown that the framework can be mechanised by implementing it using the Event-B framework, but this can also be achieved with other tool-equipped frameworks.
We used the Event-B implementation for experiments that further confirm the effectiveness of the proposed approach. We design a tool that generates for a given application, the Event-B part to be used to instantiate the generic framework; thus the process is fully automatised.

Observing that IoT-based applications are a subset of cyber-physical systems, which are mostly characterised by their heterogeneous features, the method proposed here can be generalised to these heterogeneous systems.
We conjecture that it will be of a great interest to connect our framework with existing DSLs which enable one to describe IoT systems; their descriptions will thus benefit from the formal modelling and the rigorous analysis of the designed systems prior to implementation.
We have already identified such DSLs for further investigation:
openIoT \cite{openIoT2014}, 
SDL-IoT \cite{SDLIoT2015-oatao15419},
ide4dsl \cite{ide4dsl2015-7160420},
UML4IoT \cite{THRAMBOULIDIS2016259},
SysML4IoT \cite{design_analysis_iot_2016,mbaVerifFGCS2019},
OpenHAB\footnote{\url{https://www.openhab.org/docs/}}.
 
Moreover,  we suggest to discover at least a part of the physical architecture to be controlled, using for instance a  software probe to be deployed on the dedicated network of the considered IoT system. The interest of doing like this is to ensure that the physical part description is as faithful as possible. Conversely, the physical part can be built once its abstract formal model is well-analysed and trustworthy. Finally both the abstract model of the physical part and its physical implementation can coexist during the live of the IoT system, both interacting with the sensors and actuators environment;  the abstract model (extended as necessary)  behaving then as the digital twin of the real system and enabling to check and monitor it. Such interaction between models of different abstraction levels is  planned for the future work. 
{\small \small
\bibliographystyle{plain}
}
\bibliography{\repBIBLIO/biblioIOT} 
 
\end{document}